\documentclass[]{aa} 

\usepackage{graphicx}
\usepackage{txfonts}
\usepackage[LGR,T1]{fontenc}
\usepackage{xcolor}

\newcommand{\textgreek}[1]{\begingroup\fontencoding{LGR}\selectfont#1\endgroup}

\usepackage[nolist]{acronym}
\begin{acronym}

\acro{SNR}{Supernova Remnant}
\acrodefplural{SNR}{Supernova Remnants}
\acro{WR}{Wolf-Rayet}
\end{acronym}

\begin{document}

      \title{MeerKAT reveals a ghostly thermal radio ring towards the Galactic Centre}


   \author{C. Bordiu \inst{1} \and
          M. D. Filipovi\'c\inst{2} \and
          G. Umana\inst{1} \and
          W. D. Cotton\inst{3,4} \and
          C. Buemi \inst{1} \and 
          F. Bufano \inst{1} \and
          F. Camilo \inst{4} \and  
          F. Cavallaro \inst{1} \and 
          L. Cerrigone \inst{3,5}\and 
          S. Dai \inst{2} \and 
          A. M. Hopkins \inst{6} \and 
          A. Ingallinera \inst{1} \and 
          T. Jarrett \inst{2,7}\thanks{Deceased on 3 July 2024} \and 
          B. Koribalski \inst{2,8} \and 
          S. Lazarevi\'c \inst{2,8,9} \and 
          P. Leto \inst{1} \and 
          S. Loru \inst{1} \and 
          P. Lundqvist \inst{10} \and 
          J. Mackey \inst{11} \and 
          R. P. Norris \inst{2,8} \and 
          J. Payne \inst{2} \and 
          G. Rowell \inst{12} \and 
          S. Riggi \inst{1} \and 
          J. R. Rizzo \inst{13} \and 
          A.C. Ruggeri \inst{1} \and 
          S. Shabala \inst{14} \and 
          Z. J. Smeaton \inst{2} \and 
          C. Trigilio \inst{1} \and  
          V. Velovi\'c \inst{2}
          }

   \institute{INAF–Osservatorio Astrofisico di Catania, Via Santa Sofia 78, I-95123 Catania, Italy\\
              \email{cristobal.bordiu@inaf.it}
        \and
            Western Sydney University, Locked Bag 1797, Penrith South DC, NSW 2751, Australia
        \and
            National Radio Astronomy Observatory, 520 Edgemont Road, Charlottesville, VA 22903, USA 
        \and
            South African Radio Astronomy Observatory (SARAO), 2 Fir Street, Black River Park, Observatory, Cape Town 7925, South Africa
        \and
            Joint ALMA Observatory, Alonso de C\'ordova 3107, Vitacura, Santiago, 7630355 Chile
        \and
            School of Mathematical and Physical Sciences, 12 Wally's Walk, Macquarie University, NSW 2109, Australia
        \and
            Department of Astronomy, University of Cape Town, Private Bag X3, Rondebosch 7701, South Africa
        \and
            CSIRO Space and Astronomy, Australia Telescope National Facility, PO Box 76, Epping, NSW 1710, Australia
        \and
            Astronomical Observatory, Volgina 7, 11060 Belgrade, Serbia
        \and
            Department of Astronomy, Stockholm University, The Oskar Klein Centre, AlbaNova, SE-106 91 Stockholm, Sweden
        \and
            Dublin Institute for Advanced Studies, Astronomy \& Astrophysics Section, DIAS Dunsink Observatory, Dublin, D15 XR2R, Ireland
        \and
            School of Physics, Chemistry, and Earth Sciences, The University of Adelaide, Adelaide 5005, Australia
        \and
            ISDEFE, Beatriz de Bobadilla, 3, E-28040, Madrid, Spain
        \and
            School of Natural Sciences, Private Bag 37, University of Tasmania, Hobart 7001, Australia
             }

   \date{Received ...; accepted ...}

 
  \abstract{
  We present the serendipitous discovery of a new radio-continuum ring-like object nicknamed Kýklos (J1802--3353), with MeerKAT UHF and L-band observations. The radio ring, which resembles the recently discovered odd radio circles (ORCs), has a diameter of $\sim$80\arcsec\ and is located just $\sim$6$^\circ$  from the Galactic plane.  However, Kýklos exhibits an atypical thermal radio-continuum spectrum ($\alpha=-0.1\pm0.3$), which led us to explore different possible formation scenarios. We concluded that a circumstellar shell around an evolved massive star, possibly a Wolf-Rayet, is the most convincing explanation with the present data.}


   \maketitle
%

\section{Introduction}

It is well established that observations in the radio domain are essential to investigate the late stages of stellar evolution. In particular, the radio continuum at centimetre wavelengths traces thermal emission from both the stellar winds and the abundant circumstellar ionised gas, sometimes with hints of non-thermal processes such as wind-wind collisions in binary systems. In recent decades, radio observations of planetary nebulae \citep[PNe,][]{2011MNRAS.412..223B,2016MNRAS.463..723I}, luminous blue variables \citep[LBVs,][]{2002MNRAS.330...63D, 2005A&A...437L...1U, 2010ApJ...721.1404B, 2011ApJ...739L..11U, 2012MNRAS.427.2975U}, and \ac{WR} stars \citep{1995ApJ...439..637G, 2004AJ....127.2885C, 2008RMxAC..33..142C} have built a precise portrait of these sources and their immediate surroundings, delivering for many of them accurate estimates of the total ionised gas budget.

Now, with the  Square Kilometre Array (SKA) precursors producing their first scientific results, the study of evolved stars in the radio domain benefits from an unprecedented leap forward in terms of sensitivity. Facilities such as the Australian Square Kilometre Array Pathfinder \citep[ASKAP,][]{2008ExA....22..151J, 2016PASA...33...42M,2021PASA...38....9H} and MeerKAT \citep{2009IEEEP..97.1522J,meerKAT} are offering crucial insights into the radio signature of the late evolutionary stages across a broad range of stellar masses. More importantly, these instruments have the potential to serve as discovery tools, revealing the faintest, and thus previously undetected, circumstellar structures, similar to the role of \textit{Spitzer} in the infrared \citep{2010AJ....139.1542M, 2010MNRAS.405.1047G, 2010AJ....139.2330W}.

Ongoing wide area radio continuum surveys performed with these instruments, such as the ASKAP Evolutionary Map of the Universe \citep[EMU,][]{2011PASA...28..215N} and the SARAO MeerKAT Galactic Plane Survey \citep[SMGPS,][]{10.1093/mnras/stae1166}, are revealing a significant number of low surface brightness ring-like radio sources, often associated with the late phases of stellar evolution: PNe, WR and LBV shells, or supernova remnants (SNRs). A handful of these sources, however, have been identified as a new class of astronomical object, dubbed odd radio circles (ORCs). First discovered in the EMU ASKAP survey \citep{2021PASA...38....3N}, ORCs have sizes of the order of $\sim$1\arcmin, flux densities at 1\,GHz of the order of 1--5\,mJy, and  steep non-thermal spectral indices ($\alpha<-0.4$), and are sometimes associated with galaxies at $z \sim$ 0.5. ORCs are only observed in the radio band, without obvious counterparts at other wavelengths. Because of their circular, ring-like morphologies, some authors \citep{2022MNRAS.512..265F, 2023MNRAS.526.6214S} propose that some of them are missing (inter)Galactic SNRs. However, the term ORC is now generally reserved for those with a galaxy near their geometric centre, suggesting that the ring of emission is generated by a shock interacting with previously quiescent electrons \citep{2023ApJ...945...74D,2024MNRAS.528.3854Y,2024PASA...41...24S, 2024Natur.625..459C}. The origin of these structures remains elusive and continues to be a topic of debate.

\begin{figure*}[hbt!]
\centering
    \includegraphics[width=0.9\textwidth]{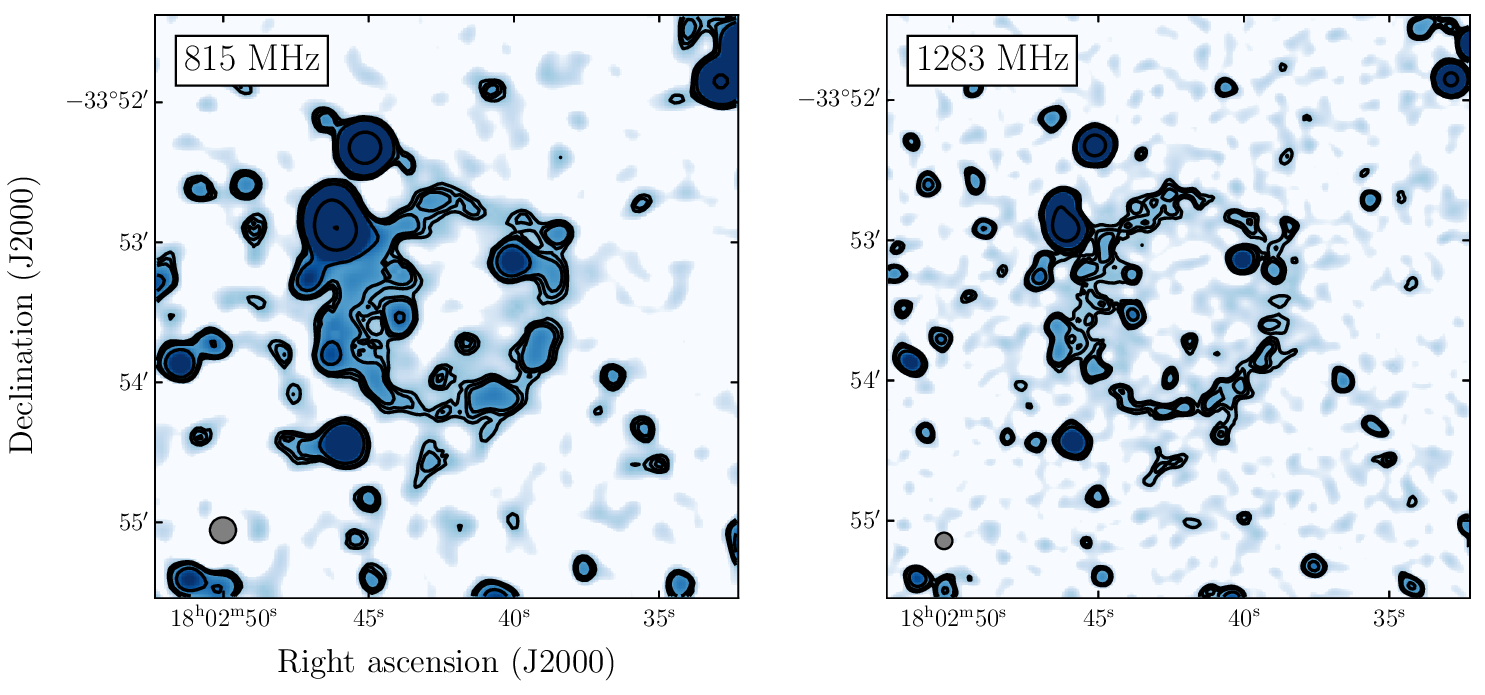}
    \caption{MeerKAT continuum images of Kýklos (J1802--3353) in UHF (left) and L-band (right), at reference frequencies of 815 and 1283\,MHz. The filled grey circles in the bottom left corner represent the native MeerKAT synthesised beam. The contours are at rms\,$\times$\,[3, 4, 5, 10, 100, 500], with rms = 6.9 and 3.6 $\mu$Jy beam$^{-1}$ in UHF and L-band, respectively.}
    \label{fig:radio-maps}
\end{figure*}

Here we report the serendipitous discovery of J1802--3353, dubbed Kýklos (from the  Greek \textgreek{κύκλος} for  circle) a new ORC-like structure just $\sim$6$^\circ$   from the Galactic plane and close (in projection) to the Galactic centre (Fig.~\ref{fig:radio-maps}). 

\section{Observations and data reduction}

The observations were made on 4 and 5 November~2022 on the MeerKAT array in L-band (856--1712\,MHz) and UHF (544--1088\,MHz) under project code SSV-20221103-SA-01. Each session included an on-source exposure of 5.5~hours spread over 6.5~hours; the observations used 62 and 61 antennas for L-band and UHF, respectively. All combinations of the two linearly polarised feeds were recorded and the bandpasses were divided into 4096 channels with an integration time of 8~seconds. The pointing position for J1802--3353 was RA(J2000)\,=\,18 02 42.91, Dec(J2000)\,=\,$-$33 53 20.2. The sources J1331+3030 and J1939--6342 were used as flux density, delay, bandpass and polarisation calibrators, with J1830--3602 as the gain calibrator. Calibration followed the procedures in \cite{Cotton2024}. The data were imaged using task MFImage \citep{Cotton2018} in the Obit package \citep{Cotton08}. 
The imaging used a Briggs Robust factor of --1.5 (AIPS/Obit usage), resulting in synthesised beams of 7.2\arcsec$\times$7.1\arcsec\ with position angle 89$^\circ$  at L-band and 11.2\arcsec$\times$11.0\arcsec\ with position angle 89$^\circ$ at UHF band. Two iterations of phase self-calibration were used in each band. 
The off-source rms of the L-band image is 3.6\,$\mu$Jy\,beam$^{-1}$ and at UHF 6.9\,$\mu$Jy\,beam$^{-1}$.

\section{Kýklos's low surface brightness ring}
\label{sec:description}

Kýklos appears as a faint ring-like structure in the UHF and L-band MeerKAT total intensity maps (see Fig.~\ref{fig:radio-maps}). The ring is clumpy and nearly circular, with an approximate diameter of $\sim$80\arcsec\ and J2000 equatorial coordinates of the centroid of $(\alpha,\delta) = (270\fdg67604, -33\fdg89124)$.
No clear central source is detected in either of the   bands. The ring is thin, with a thickness of $\sim$6\arcsec,\ as measured in the L-band image, and slightly thickens towards the NE edge. This effect is particularly evident in the UHF band image.

Numerous compact radio sources, most likely unrelated and possibly extragalactic, are spread across the observed field. The most prominent among these is a bright, slightly elongated source that partially overlaps the ring on its NE side. At least five other obvious point-like sources, easily distinguishable in the L-band image, are located within the ring.

\begin{figure*}
\centering
    \includegraphics[width=0.95\textwidth]{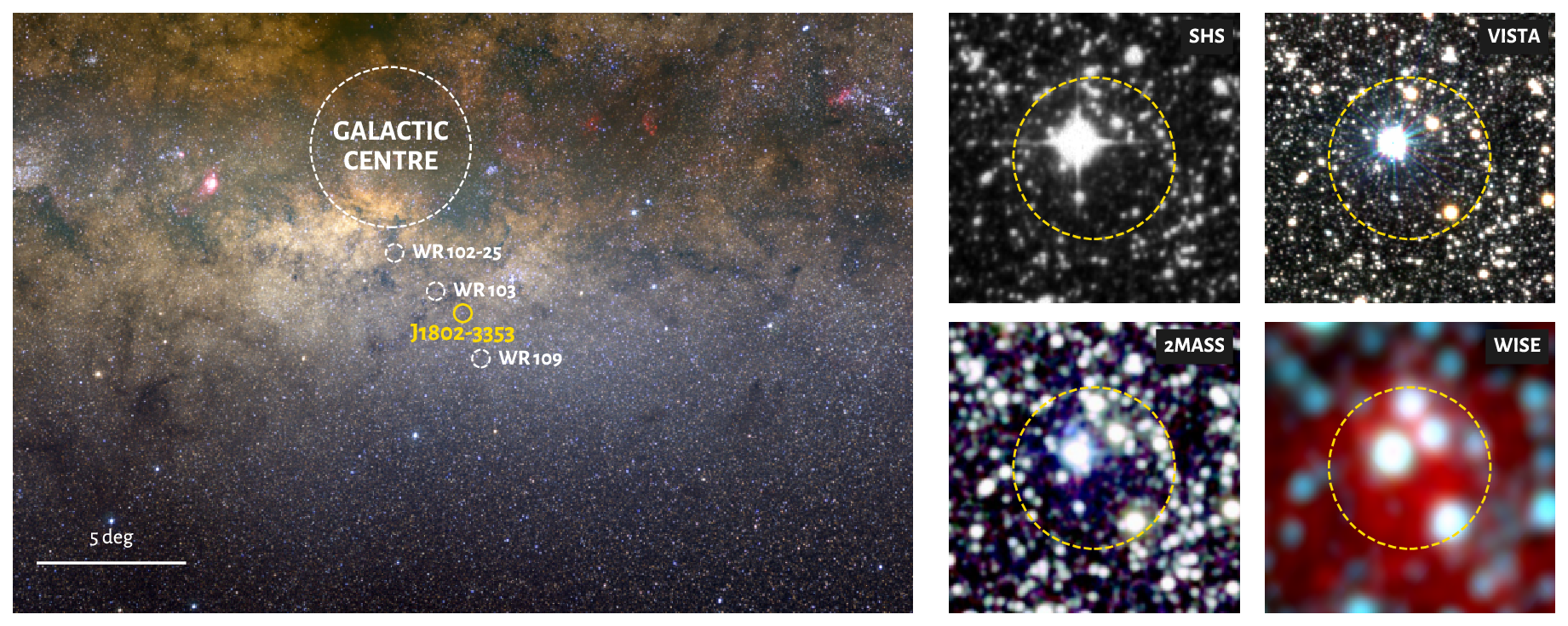}
    \caption{Multiwavelength view of Kýklos. Left: RGB composite optical image showing the location of Kýklos radio shell (yellow circle, not to scale) relative to the Galactic plane and the Galactic centre. Nearby catalogued Wolf-Rayet stars WR~102--25, WR~103, and WR~109 are indicated as dashed white circles. Image credit: A. Mellinger, Central Michigan University \citep{2009PASP..121.1180M}. Right: Zoomed-in view of the field of Kýklos, as seen in the SuperCOSMOS H$\alpha$ survey, the VISTA VVV survey (J/Y/Z), 2MASS (J/H/K), and WISE (W1/W2/W4). The yellow dashed circle represents the size of the radio shell (80\arcsec).}
    \label{fig:fig-context}
\end{figure*}

Unlike known ORCs, Kýklos lies just $\sim$6$^\circ$ below the Galactic plane, and in projection very close to the Galactic Centre (Fig.~\ref{fig:fig-context}). Quite remarkably, the position of Kýklos places it in an almost straight line with three catalogued \ac{WR} stars, namely WR\,102--25, WR\,103, and WR\,109.

We searched for a counterpart of Kýklos in multiwavelength archives. No associated X-ray emission is detected in the eROSITA-DE DR1 data. At optical and  near- and mid-infrared wavelengths, the field is clearly dominated by the bright blue star HD\,164455 (Fig.~\ref{fig:fig-context}, right panels). No H$\alpha$ emission from the ring is detected in the SuperCOSMOS H$\alpha$ survey (SHS), nor in VPHAS+, and no obvious infrared counterpart is observed up to 22 $\mu$m. The area is dotted by multiple infrared sources, the majority of which are of unknown nature. The WISE~W4 image reveals a diffuse glow extending well beyond the edges of the radio ring, without a clear association. This is likely the combined effect of the (broad) point spread function from the brightest point-like sources and the background structure. Unfortunately, the coverage of the region in the mid- and far-infrared range is very limited, which prevents us from assessing the existence of a cold dust component. Neither the \textit{Spitzer} nor the \textit{Herschel} telescopes have imaged the field;   only AKARI/FIS imagery is available in the range 65--160\,$\mu$m, which provides insufficient sensitivity and angular resolution for the study of J1802--3353.

\subsection{Kýklos's flux density}

Kýklos is a low-brightness source located in a relatively crowded region. Consequently, obtaining a reliable flux density is a challenge. One must be especially cautious to ensure that no nearby or overlapping compact sources contaminate the measured flux as this would bias the spectral index analysis. Likewise, a proper determination of the local background becomes critical at the observed frequencies.

To measure the integrated flux density of Kýklos, we first convolved the L-band map to the slightly larger beam of the UHF band image and re-gridded it to the same pixel size. Then, we employed standard aperture photometry, selecting a polygon region (equal for both maps) encompassing all the faint emission from the ring, and carefully excluding any obvious unrelated compact sources. We estimated the background within an annulus-like region closely surrounding Kýklos, again excluding all obvious nearby sources. This method yields a flux density of 0.57$\pm$0.05\,mJy at 815\,MHz, and 0.54$\pm$0.04\,mJy at 1283\,MHz. 

The ring's appearance resembles a limb-brightened shell, with the interior devoid of detectable emission. In an optically thin and  perfectly uniform spherical shell of inner radius $R_1\sim$37\arcsec\ and outer radius $R_2\sim$43\arcsec, the ratio of the line-of-sight path length through the shell centre, $L_\mathrm{centre} = 2(R_2-R_1)$, to the maximum path length through the limb, $L_\mathrm{limb} = 2\sqrt(R_2^2-R_1^2)$, is proportional to the column density ratio, and equals the centre-to-limb surface brightness ratio. For an average L-band limb brightness $S_\mathrm{limb} \sim$10 $\mu$Jy beam$^{-1}$, the expected brightness at the shell centre would be $S_\mathrm{centre}\sim3$ $\mu$Jy beam$^{-1}$, comparable to the map rms.

\subsection{Kýklos's spectral index}

The integrated spectral index resulting from the derived flux densities is $\alpha$\,=\,--0.12$\pm$0.56, compatible with thermal (bremsstrahlung) emission and consistent with the canonical value of an optically thin H\textsc{ii} region ($\alpha$\,=\,--0.1) \citep{2016era..book.....C,book2}. However, the large uncertainty is problematic and calls into question any strong quantitative claims about the nature of the emission. This uncertainty is a consequence of the source's faintness and the limited frequency coverage of the observations ($\nu_2$/$\nu_1$\,$\sim$\,1.5). 

\begin{figure}[hbt!]
\centering
\includegraphics[width=\columnwidth]{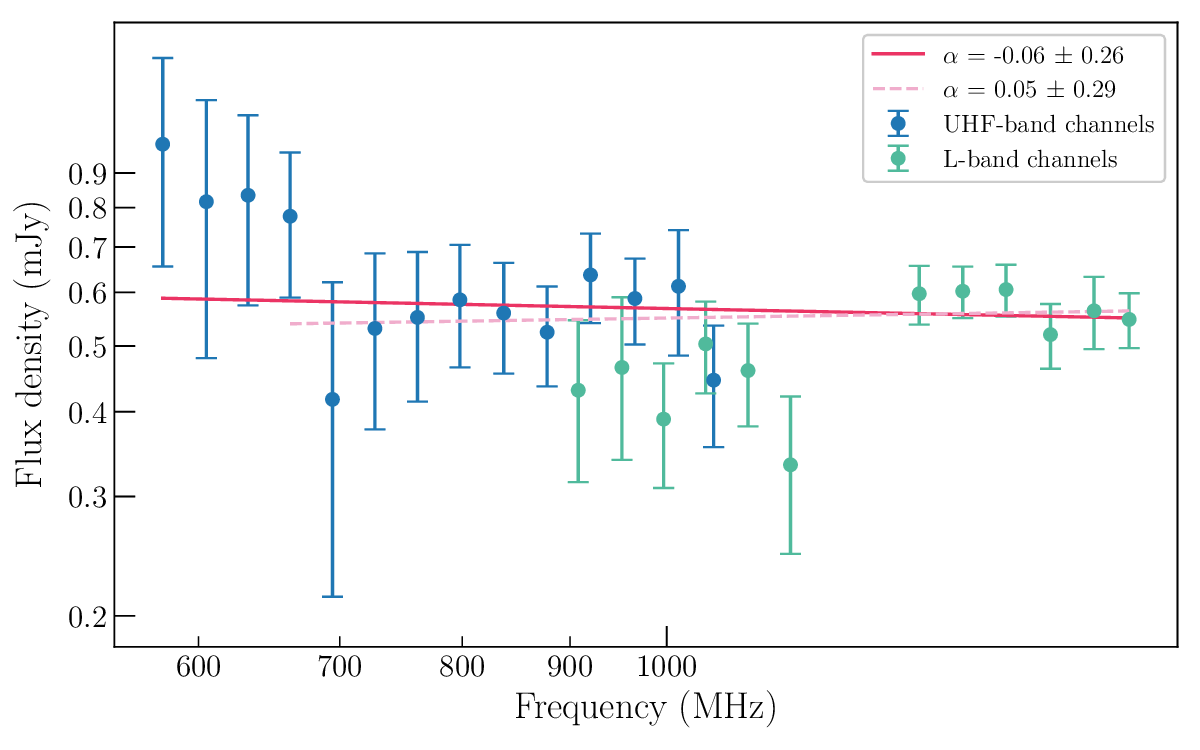}
\caption{MeerKAT UHF and L-band spectrum of Kýklos. The red line represents the power law fit to the data, corresponding to a spectral index $\alpha=-0.06\pm0.26$. The dashed line shows the fit when excluding the first channels in which the source is not clearly detected (see text).}
\label{fig:spix-fit}
\end{figure}

To mitigate this uncertainty and confirm the spectral index of the ring, we resorted to the analysis of the full spectral cubes, aiming for a more robust fit. We have 26 frequency channels extending from $\sim$560 to $\sim$1680\,MHz: 14 in UHF band and 12 in L-band (channels~7 and 8 are flagged due to radio frequency interference), with some overlapping between the upper UHF band and the lower L-band. We measured the flux density in each channel within the same region employed for the total intensity maps. Then we fitted a simple power law model to the data using the \texttt{curve\_fit} method available in the \texttt{scipy} package,\footnote{\url{https://docs.scipy.org/doc/scipy/reference/generated/scipy.optimize.curve_fit.html}} as shown in Fig.~\ref{fig:spix-fit}. The resulting slope is $\alpha$\,=\,--0.06$\pm$0.26, slightly better constrained and compatible with the value derived from the total intensity maps. However, a closer inspection of the  individual frequency channels reveals that the ring is not discernible in the (noisier) lower UHF band channels, which correspond to the highest flux densities,  and have the largest associated errors. This probably indicates that we are just measuring the background. On the contrary, the ring is more clearly visible in the upper L-band, which again is suggestive of a thermal spectrum. When we exclude the initial UHF band channels from the fit, the slope becomes positive ($\alpha$\,=\,0.05$\pm$0.29).

\begin{figure}[hbt!]
\centering
\includegraphics[width=\columnwidth]{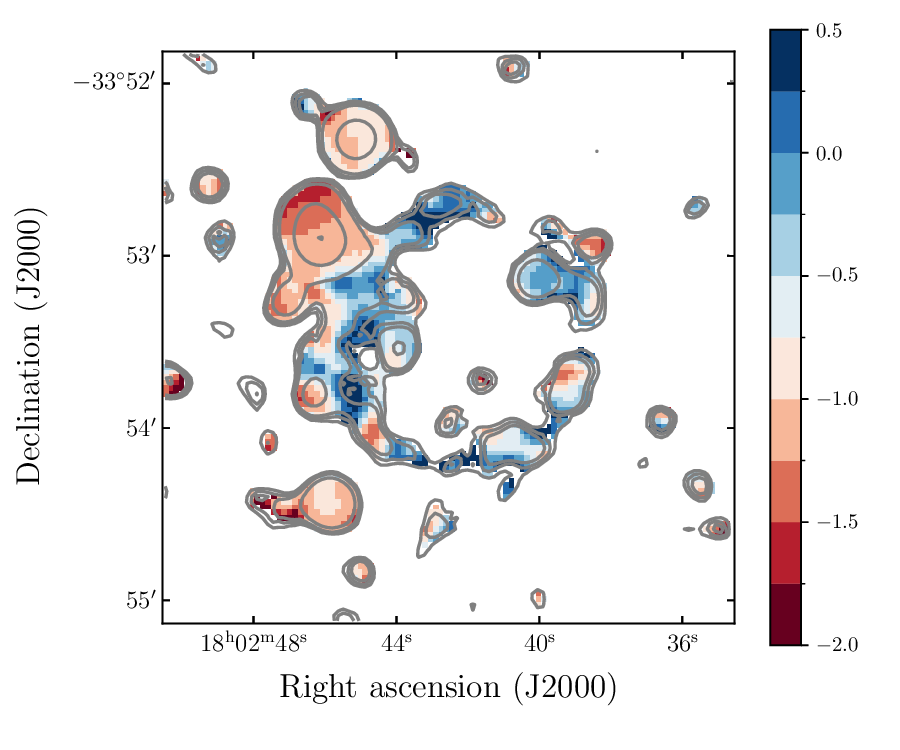}
\caption{MeerKAT spectral index map of the Kýklos ring, made using UHF and L-band images (0.815 and 1.283 GHz, respectively). The UHF band contours (as in Fig. \ref{fig:radio-maps}) are superimposed in grey. Pixels with brightness below 3$\sigma$ in any of the bands were masked. }
\label{fig:spix-map}
\end{figure}

Finally, we also produced a spectral index map, presented in Fig.~\ref{fig:spix-map}.  Even though the associated error per pixel is high ($\sim$0.7), the map is useful for a qualitative assessment: it is immediately apparent that the ring has a much flatter spectrum compared to other prominent sources in the field. These sources display very steep, negative spectral indices characteristic of radio galaxies, whereas Kýklos has an average spectral index value of $\sim$--0.1, in good agreement with previous estimates. The map reveals some fluctuations within the ring, but it is difficult to assess whether they are meaningful (stemming from changes in the local physical conditions) or arise from the large uncertainty.

All things considered, we claim that the emission from the shell Kýklos is mostly thermal, even if our ability to put tight constraints on the spectral index is limited with the present data. For the remainder of the analysis, we conservatively adopt a spectral index $\alpha$\,=\,--0.1$\pm$0.3.

\subsection{Kýklos's emission measure}
\label{subsec:em-meas}

If the emission from Kýklos is indeed free-free, we can compute its emission measure (EM), defined as the line-of-sight integral of the electron density ($n_\mathrm{e}$) squared:

\begin{equation}
    \text{EM} = \int_0^{\infty} n_\mathrm{e}^2 \, dl
.\end{equation}

\noindent Following \cite{1967ApJ...147..471M}, in an optically thin isothermal ionised gas, the continuum brightness temperature $T_\mathrm{C}$ is related to the EM, such that

\begin{equation}
    T_\mathrm{C} = T_\mathrm{e} \left(1 - e^{-\tau_C}\right) \simeq T_\mathrm{e} \tau_\mathrm{C}
,\end{equation}

\noindent with $T_\mathrm{e}$ the electron temperature and $\tau_\mathrm{C}$ the optical thickness of the continuum, given by

\begin{equation}
\tau_\mathrm{C} = 8.235 \times 10^{-2}\, T_\mathrm{e}^{-1.35} \left( \frac{\nu}{\text{GHz}} \right)^{-2.1} \text{EM}
.\end{equation}

Using the Rayleigh-Jeans approximation to convert the average surface brightness to brightness temperature and adopting a typical $T_\mathrm{e}$ = $10^4$ K, we obtain an EM of $\sim$ 60 pc cm$^{-6}$, a moderate value that indicates low-density ionised gas. The EM is in turn related to the H$\alpha$ intensity \citep{1998PASA...15..111V} through the expression

\begin{equation}
    I_\mathrm{H\alpha} = 9.41\times10^{-8}\,T_{\mathrm{e}4}^{-1.017}10^{-0.029/T_{\mathrm{e}4}}\, \text{EM}
,\end{equation}

\noindent where $T_{\mathrm{e}4}$ is the electron temperature in units of $10^4$ K, and $I_\mathrm{H\alpha}$ denotes the H$\alpha$ intensity in units of erg cm$^{-2}$ s$^{-1}$ sr$^{-1}$. For the derived EM we obtain an intensity $I_\mathrm{H\alpha}$ $\sim$22 rayleigh.\footnote{One rayleigh (R) is equivalent to 2.41$\times$ 10$^{-7}$ erg cm$^{-2}$ s$^{-1}$ sr$^{-1}$ at the H$\alpha$ wavelength.} This value is above the sensitivity limit of most H$\alpha$ surveys (e.g. 5 rayleigh in SHS; \citealt{2005MNRAS.362..689P}). Therefore, the non-detection of an H$\alpha$ counterpart is surprising and can likely be attributed to either interstellar or circumstellar extinction.

\section{The nature of Kýklos}

From a purely morphological perspective, the Kýklos ring resembles an ORC: a faint radio ring of $\sim$80\arcsec, with no known counterpart at any other wavelength. However, it exhibits several peculiarities that make it stand out among the ORC population, the main characteristics of which are listed in Table~\ref{tab:ORC-properties}. In particular, 1) it is located at a much lower Galactic latitude ($\sim$--6$^\circ$); 2) it is almost one order of magnitude fainter at $\sim$1\,GHz; and 3) it has a much flatter spectral index, possibly tracing thermal emission ($\alpha\approx$\,--0.1). These characteristics do not align with the preferred explanations for the ORC phenomenon, making us reevaluate some possibilities previously discussed by \citet{2021PASA...38....3N}.

\begin{table*}
\caption{Main properties of ORCs and ORC candidates.}\label{tab:ORC-properties}
\begin{tabular}{@{}llccccccll@{}}
\hline
\noalign{\smallskip}
ORC ID & Source & Diam. & $l$ & $b$ & $F_\mathrm{1\,GHz}$ & $\alpha$ & Host & Instrument & Refs \\ 
        &       &   (arcsec)        &   (deg)   & (deg) & (mJy) &   &  galaxy?   &   & \\
\hline
\noalign{\smallskip}

ORC 1 & ORC J2103--6200  & 80  & 333.41592 & --39.00906 & 3.9   & --1.2$\pm$0.1  & \checkmark & ASKAP   & 1 \\
ORC 2 & ORC J2058--5736a & 80  & 339.08813 & --39.52277 & 7.0   & --0.8$\pm$0.1 &   ---   & ASKAP   & 1 \\
ORC 3 & ORC J2058--5736b & 80  & 339.08147 & --39.55247 & 1.9   & --0.5$\pm$0.2  &   ---   & ASKAP   & 1 \\
ORC 4 & ORC J1656+2726  & 90  & 044.35860 & +49.36566 & 9.4   & --0.9$\pm$0.2  & \checkmark & GMRT    & 1           \\
ORC 5 & ORC J0102--2450  & 70  & 344.00735 & +24.60441 & 3.2   & --0.8$\pm$0.2  & \checkmark & ASKAP   & 2       \\
 & ORC J0624--6948  & 196 & 280.16797  & --27.66217 & 11.2\tablefootmark{a} & --0.4$\pm$0.1  &  ---    & ASKAP   & 3   \\
 & ORC J1027--4422  & 90  & 277.53776 &  +11.26383 & $\sim$1.0 & $>-1.9$ & --- & MeerKAT & 4 \\
 \hline
\noalign{\smallskip}
      & J1802--3353  & 80  & 357.55482 & --5.64533 & 0.5   & --0.1$\pm$0.3  &  ?    & MeerKAT & 5 \\
\hline
\end{tabular}
\tablefoot{\tablefoottext{a}{Scaled from the 888 MHz flux density reported in \citet{2022MNRAS.512..265F}.}}
\tablebib{
(1)~\citet{2021PASA...38....3N};
(2)~\citet{2021MNRAS.505L..11K};
(3)~\citet{2022MNRAS.512..265F};
(4)~\citet{2024MNRAS.531.3357K};
(5)~this work.
}
\end{table*}

\subsection{Kýklos as an extragalactic source}

The archetypal ORCs (ORCs 1, 4, and 5, \citealt{2021PASA...38....3N}) have a prominent optical galaxy ($z\sim$ 0.5) located near the geometric centre. Such an occurrence is unlikely to happen by chance, strongly supporting ORC's extragalactic origin, in which these structures would trace shocks from starburst events \citep{2021PASA...38....3N, 2024Natur.625..459C}, supermassive black hole mergers \citep{2022MNRAS.513.1300N}, galaxy mergers \citep{2023ApJ...945...74D}, or shock excitation of a relic lobe \citep{2024PASA...41...24S}, among other proposed hypotheses. We therefore searched for optical galaxies potentially associated with Kýklos.

\begin{figure}[hbt!]
\centering
\includegraphics[width=\columnwidth]{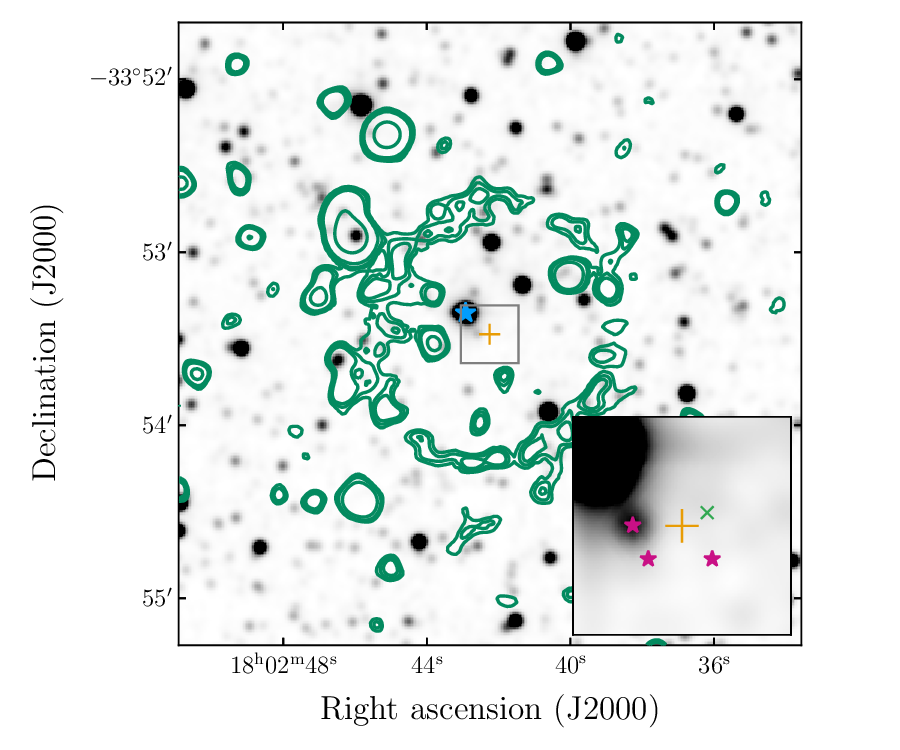}
\caption{MeerKAT L-band contours of Kýklos (as in Fig.~\ref{fig:radio-maps}) superimposed onto an optical DSS2 red image of the field. The orange plus marker indicates the centroid of the shell, and the blue star marker represents HD~164455. The inset shows a zoomed-in image of the central 20\arcsec$\times$20\arcsec\ region, with the positions of three \textit{Gaia} stars and a galaxy candidate (see text) indicated by magenta stars and a green cross, respectively.}
\label{fig:fig-crossmatch}
\end{figure}

Unfortunately, the Dark Energy Survey \citep[DES;][]{2018ApJS..239...18A}, in which the host galaxies of ORCs 1, 4, and 5 were identified, does not cover the position of Kýklos. Likewise, the NASA/IPAC Extragalactic database (NED) does not catalogue any confirmed galaxies in the region.  We thus resorted to optical data from \textit{Gaia}. The \textit{Gaia} archive lists a total of 428 sources that fall within the boundaries of the shell. The majority of these sources are either unclassified or classified with a 100$\%$ probability of being stars, with only two labelled as potential galaxy or quasar candidates (\textit{Gaia} DR3 4042247335830104960, \textit{Gaia} DR3 4042247331685419136). While \textit{Gaia} DR3 4042247331685419136 is almost at the SW edge of the shell, \textit{Gaia} DR3 4042247335830104960 is just $\sim$3\arcsec\ from the centroid, which makes an association plausible. Its relative location is represented in Fig.~\ref{fig:fig-crossmatch} with a green marker. \textit{Gaia} assigns this source a probability of $\sim$99\% of being a galaxy, and it has a similar brightness ($M_{G}$\,=\,19.6695\,mag) to that of other ORC host galaxies. Therefore, we cannot dismiss the possibility of Kýklos being an extragalactic source (i.e. a bona fide ORC), and yet explaining its flat spectrum would be challenging and would require a revision of current formation models.

\subsection{Kýklos as a Galactic supernova remnant}

In general, some ORC candidates resemble SNRs due to the absence of optical/IR counterparts and their steep non-thermal spectral indices; however, their location at high Galactic latitudes conflicts with the distribution of Galactic SNRs \citep{2021PASA...38....3N}. In this context, \cite{2022MNRAS.512..265F} proposed ORC~J0624--6948 as an (inter)Galactic SNR originating from a type~Ia SN in the outskirts of the Large Magellanic Cloud.

The question arises of whether Kýklos could be a shell-type Galactic SNR. This interpretation is quite problematic, even if its location and morphology may suggest otherwise. First, we cannot completely rule out the existence of a dusty counterpart due to the poor infrared coverage, especially in the far-IR range (key to detecting the emission from warm dust). Second, and more importantly, its spectral index of $\alpha$\,=\,--0.1$\pm$0.3, significantly deviates from the typical value of $\alpha$\,=\,--0.51$\pm$0.01 \citep{2023ApJS..265...53R}, even when considering the large uncertainty. Similarly, the absence of detectable X-ray emission in eROSITA further weakens the SNR scenario. Finally, we note that SIMBAD lists four pulsars within 30\arcmin of the centroid;  the nearest one is at $\sim7$\arcmin, but without proper motion information, a possible association cannot be established.

\subsection{Kýklos as a planetary nebula}
\label{sec:pn}

Planetary nebulae are mainly detected in the optical due to their intense H$\alpha$ emission from ionised gas \citep{2014MNRAS.443.3388S}. However, significant extinction near the Galactic plane can sometimes render optical detection unfeasible, leaving the radio window as the only alternative for the identification and study of PNe \citep{2011MNRAS.412..223B}. Notably, numerous PNe manifest as rings and shells at radio wavelengths, with flat thermal spectra and typically lacking a central point source \citep{2016MNRAS.463..723I}. These properties match perfectly the observed characteristics of Kýklos.

Even so, there is a counter-argument that challenges the interpretation of Kýklos as a `missing' PN. Compared to the known PNe in the HASH PNe database \citep[v4.6, 3309 entries,][]{2016JPhCS.728c2008P}, Kýklos exceeds the 85${\rm th}$ percentile in angular size. This implies that the shell is either relatively close or intrinsically large (corresponding to an old, highly evolved PN), or both. 

If Kýklos is nearby, it is reasonable to expect lower extinction, especially given its location $\sim$6$^\circ$  from the Galactic plane. The \cite{2006A&A...453..635M} Galactic extinction model gives, for the nearest lines of sight, infrared absorptions $A_{K_s}$\,<\,0.2 mag up to distances of 8--10 kpc. This translates to $A_{V}$\,<\,2 mag, which are low extinction values consistent with the observed crowdedness of the field in optical and infrared imagery (see Fig.~\ref{fig:fig-context}). Clear hints of the ring should then be visible at these wavelengths, and yet we see none, considering that the diffuse WISE W4 emission seems unrelated as discussed in Sect. \ref{sec:description}. 

The H$\alpha$ intensity predicted in Sect. \ref{subsec:em-meas} appears to be inconsistent with the derived optical extinction. However, we note that 1) the \cite{2006A&A...453..635M} model has a relatively coarse spatial resolution (15\arcmin$\times$15\arcmin) and might not be accurate in the exact direction of the ring; 2) we cannot disregard an extinction excess caused by an unseen circumstellar dust component; and 3) evolutionary factors such as a low hydrogen abundance could also play a role in reducing the H$\alpha$ signature.

Conversely, if we apply the empirical $\Sigma-D$ relation for PNe \citep{2009A&A...495..537U,2009A&A...503..855V} extrapolating the flux density at 5~GHz with $\alpha$\,=\,--0.1, the resulting physical diameter is $\sim$3.5\,pc at 9\,kpc. Even if theoretically possible, such a diameter would place Kýklos among the largest PNe known \citep{2009MNRAS.399..769F,2016MNRAS.455.1459F}, far surpassing the typical size of PNe, which rarely exceeds $\sim$1\,pc. Nonetheless, even in a highly evolved PN scenario, traces of the shell might still be detectable at shorter wavelengths. The dust would have been destroyed long ago by the increasing UV flux of the central source, but the optical and infrared emission would now be dominated by highly ionised gas lines \citep{2011ApJ...741....4F,2016MNRAS.463..723I}. Given these considerations, the PN nature of Kýklos appears unlikely.

\subsection{Kýklos as ejecta from an old nova outburst}

In a more speculative scenario, Kýklos could be the old remnant of a classical nova outburst. Radio continuum emission from nova shells is a mixture of thermal and non-thermal contributions \citep{1979AJ.....84.1619H,2021ApJS..257...49C}, with some sources having spectra completely dominated by thermal bremsstrahlung ($\alpha\sim-0.1$) as the warm ionised ejecta expands \citep{2023PASA...40...25G}. This aligns well with the spectral index of Kýklos.

Eventually, the shell expansion causes the recombination rate to decrease, with the subsequent fading of the associated H$\alpha$ emission. This fading, together with extinction (as discussed for the PN scenario in Sect. \ref{sec:pn}), could explain the non-detection of an H$\alpha$ counterpart. Indeed, several systematic H$\alpha$ searches for nova shells around cataclysmic variables have largely been unfruitful \citep{2015MNRAS.449.2215S,2015MNRAS.451.2863S,2022MNRAS.510.4180S}, with only a few  ancient\, nova shells detected in the Milky Way. These results suggest long recurrence times for these outbursts (of a few kyr), and very low surface brightnesses for the associated remnants.

In this context, the nova hibernation hypothesis \citep{1986ApJ...311..163S} could explain the lack of an evident central source in Kýklos. According to this hypothesis, after an outburst, the binary system could become detached for an extended period, entering a quiescent phase with a substantial decline in luminosity until mass transfer is resumed. This scenario is also consistent with the non-detection of X-ray emission in eROSITA.

An alternative mechanism that could account for a large circumstellar ionised shell involves an obscured recurrent nova. The secular accumulation of ejecta from successive outbursts could lead to the formation of a supershell\, (or  SNR) at parsec scales, as recently observed in the yearly recurrent nova M31 2008-12a \citep{2019Natur.565..460D}.

\subsection{Kýklos as a massive star mass-loss relic}

Mass loss from early-type evolved massive stars can produce circumstellar shells of ionised material, easily detectable in the radio band. Such ring-like structures, resulting from episodic outbursts or the accumulation of material by the action of different wind regimes, have been frequently observed around many Galactic LBVs (\citealt{2002MNRAS.330...63D, 2005A&A...437L...1U, 2011ApJ...739L..11U}, Umana et al., in prep.), and also in a handful of Galactic WR stars (\citealt{2002AJ....123.3348C,2008RMxAC..33..142C}, Buemi et al., in prep.). 

Generally, these circumstellar shells present a flat radio spectrum comparable to that of Kýklos. However, they invariably exhibit a multiwavelength footprint, coexisting with dusty \citep{2010ApJ...721.1404B, 2013A&A...557A..20V, 2015A&A...578A.108V} and, in certain instances, molecular counterparts \citep{2003A&A...411..465R, 2008ApJ...681..355R, 2019MNRAS.482.1651B, 2021MNRAS.500.5500B}. In this context, the apparent absence of noticeable infrared emission would be difficult to reconcile with a  LBV scenario as efficient dust production is expected, provided a high enough mass-loss rate \citep{2011ApJ...743...73K}. On the contrary, the situation is more readily explained in the case of a WR shell: any pre-existing dust produced in a previous phase would have been progressively destroyed as the star developed its fast wind, leaving only the most distant and coldest dust, emitting at wavelengths longer than 24\,$\mu$m, for which no data is available.

To delve deeper into the hypothesis of a mass-loss relic, we must explore the question of the parent star of Kýklos. Given its position within the shell, star HD~164455 is a primary suspect. Its slightly offset location from the shell centroid is not a problem per se, as a shell expanding in an inhomogeneous medium could explain it. We note, though, that its proper motion ($\mu_{\alpha}\cos{\delta} = 0.82 \pm 0.05$\,mas\,yr$^{-1}$, $\mu_{\delta} = -5.45 \pm 0.03$\,mas\,yr$^{-1}$) would make the star move roughly towards the shell centroid, which is unlikely. Furthermore, there are several other reasons why HD~164455 is not a suitable candidate. HD~164455 is a B2III/IV star located at a distance of just 810\,pc which would suggest a shell diameter of $\sim$0.3\,pc. Being a (sub)giant star \citep{2010AN....331..349H}, its luminosity (2.6$\times$10$^3$\,L$_\odot$), mass (6.75$\pm$0.35\,M$_\odot$), and evolutionary status are incompatible with an LBV-like or WR scenario, and its mass-loss rate has   probably been too low for it to accumulate sufficient material to produce an ionised shell. 

In the vicinity of the shell centroid, there are several other optical and infrared candidates. Narrowing the search down to the central 10\arcsec, \textit{Gaia} lists three stellar sources ($p_\mathrm{star}$\,$\sim$\,99\%), whose locations are indicated in Fig.~\ref{fig:fig-crossmatch} (magenta star markers). One of them is coincident with the 2MASS source J18024263--3353280, a bluish star at $\sim$7.5\,kpc (geometric distance from \textit{Gaia} parallax, \citealt{2021AJ....161..147B}). At this distance, the ring diameter would be $\sim$3\,pc. However, no further information is available for these sources and, without spectroscopy, an unambiguous identification of the parent star is not possible. Radio observations at higher frequencies could reveal a central point-like radio source, assuming a single-star scenario where the emission originates from a thermal stellar wind ($\alpha\,$=\,0.6).

The very absence of an evident central point source in the L-band provides clues to the nature of a putative parent star. We can leverage the \cite{1975A&A....39....1P} relation to assess the  detectability of a thermal stellar wind from different types of evolved stars. The theoretical flux density scales with frequency as $F_\nu \propto \nu^{0.6}$, with the wind mass-loss rate as $\propto \dot M^{4/3}$ and with the wind velocity as $\propto v_\mathrm{exp}^{-4/3}$. Hence, at any given distance, it is easier to accomplish an L-band detection of an LBV star than a WR star, owing to the former's higher mass-loss rates and slower wind velocities. At a typical distance of 5 kpc, we would expect flux densities of the order of a few mJy for LBVs, compared to a few $\mu$Jy for WR stars. This rule-of-thumb approach suggests that if the circumstellar shell hypothesis is correct, the parent star is possibly a WR, although it could just as well be a quiescent LBV with a lower mass-loss rate. Furthermore, the relative proximity (in projection) of WR stars WR\,102-25, WR\,103, and WR\,109 (Fig. \ref{fig:fig-context}) demonstrates that finding a WR star at this Galactic latitude is plausible, even if their apparent alignment appears to be coincidental.

\section{Summary and conclusions}

We report the discovery of J1802--3353 (Kýklos), an extremely faint radio ring of $\sim$80\arcsec, in the  MeerKAT UHF and L-band images. Given its peculiarities, we   briefly explored several potential explanations for its origin, including a shock-related extragalactic structure (i.e. an ORC), a Galactic SNR, a PN, a nova remnant, and a circumstellar shell around an evolved massive star. Based on the limited data currently available, the morphological and spectral characteristics of Kýklos appear more consistent with those of a WR shell. This interpretation is further supported by the absence of a detectable central point source in the L-band image.

However, it is important to note that the limited frequency coverage of our radio observations (resulting in a loosely constrained spectral index) and the lack of infrared data in the wavelength range from 24 to 160\,$\mu$m (critical to evaluating the existence of different dust populations) greatly hamper our ability to settle the question of the nature of Kýklos. Future multiwavelength follow-up observations will be essential to fully characterise the ring and identify a possible central source, confirming the WR shell hypothesis, or perhaps revealing more intriguing scenarios. 

Regardless, the discovery of Kýklos underscores the potential of SKA precursors like MeerKAT for serendipitous findings. There remains much to be uncovered in the radio sky, and as we probe deeper into the low surface brightness universe, an increasing number of unsuspected structures will be found. The SMGPS Extended Source Catalogue \citep{Bordiu2024}, which contains thousands of unclassified radio structures, is an excellent example of what is to come.

\begin{acknowledgements}
The MeerKAT telescope is operated by the South African Radio Astronomy Observatory, which is a facility of the National Research Foundation, an agency of the Department of Science and Innovation. The National Radio Astronomy Observatory is a facility of the National
Science Foundation, operated under a cooperative agreement by Associated Universities, Inc.
JM acknowledges support from a Royal Society -- Science Foundation
Ireland University Research Fellowship (20/RS-URF-R/3712). This research made use of \texttt{hips2fits},\footnote{https://alasky.cds.unistra.fr/hips-image-services/hips2fits} a service provided by CDS. 
We thank the anonymous referee for useful comments and suggestions that greatly improved our paper.
\end{acknowledgements}

\bibliographystyle{aa}
\bibliography{references}

\end{document}